\documentclass[conference]{IEEEtran}
\IEEEoverridecommandlockouts

\usepackage{cite}
\usepackage{amsmath,amssymb,amsfonts}
\usepackage{algorithmic}
\usepackage{graphicx}
\usepackage{textcomp}
\usepackage{xcolor}
\def\BibTeX{{\rm B\kern-.05em{\sc i\kern-.025em b}\kern-.08em
    T\kern-.1667em\lower.7ex\hbox{E}\kern-.125emX}}
\usepackage{microtype}
\usepackage{caption}
\usepackage{subcaption}
\usepackage{bm}

\usepackage{graphicx}
\usepackage{filecontents} 

\begin{document}

\title{Machine Learning-Based Channel Prediction for RIS-assisted MIMO Systems With Channel Aging
\thanks{This research was supported by the Research Council of Finland (former Academy of Finland) 6G Flagship Programme (Grant Number: 346208).}}
\author{\IEEEauthorblockN{Nipuni  Ginige,  Arthur Sousa de Sena, Nurul Huda Mahmood, Nandana Rajatheva, and Matti Latva-aho}\IEEEauthorblockA{Centre for Wireless Communications,
University of Oulu,
Finland \\
\{nipuni.ginige, arthur.sena, nurulhuda.mahmood, nandana.rajatheva, matti.latva-aho\}@oulu.fi}}

\maketitle

\begin{abstract}  
Reconfigurable intelligent surfaces (RISs) have emerged as a promising technology to enhance the performance of sixth-generation (6G) and beyond communication systems. The passive nature of RISs and their large number of reflecting elements pose challenges to the channel estimation process. The associated complexity further escalates when the channel coefficients are fast-varying as in scenarios with user mobility. In this paper, we propose an extended channel estimation framework for RIS-assisted multiple-input multiple-output (MIMO) systems based on a convolutional neural network (CNN) integrated with an autoregressive (AR) predictor. The implemented framework is designed for identifying the aging pattern and predicting enhanced estimates of the wireless channels in correlated fast-fading environments. Insightful simulation results demonstrate that our proposed CNN-AR approach is robust to channel aging, exhibiting a high-precision estimation accuracy. The results also show that our approach can achieve high spectral efficiency and low pilot overhead compared to traditional methods.
 
\end{abstract}

\begin{IEEEkeywords}Reconfigurable intelligent surfaces, channel estimation, channel aging, Autoregressive processes, CNN.
\end{IEEEkeywords}

\section{Introduction}
Reconfigurable intelligent surface (RIS) has become a promising technology to enhance the performance of sixth-generation (6G) and beyond systems, offering the potential to manipulate the electromagnetic propagation environment through the configuration of nearly passive reflecting elements. An RIS comprises a large number of low-cost elements that are capable of inducing phase changes to incoming signals toward diverse objectives, including improving the quality of service (QoS) in scenarios where receivers are experiencing weak signal reception or when the direct path between the transmitter and receiver is completely blocked. The RIS technology functions as a smart reflection hub that can support massive connectivity, manage interference through strategic passive beamforming, and mitigate security vulnerabilities\cite{Liu21_RIS_survey}.

Channel estimation plays a crucial role in modern multiple-input multiple-output (MIMO) systems, being the foundation for various signal processing and precoding techniques essential to enable features such as spatial diversity and multiplexing, interference mitigation, and dynamic adaptation to fast-varying wireless environments. The accuracy of the channel state information (CSI) has, consequently, a direct impact on the system performance, spectral efficiency, and reliability. The importance of channel estimation to RIS-assisted MIMO systems is even more pivotal as, in addition to traditional active base station (BS) precoders, the passive RIS reflecting coefficients need to be carefully optimized based on the additional RIS reflected channels, which are generally challenging to acquire.

Channel estimation in RIS-assisted systems can be carried out based on either semi-passive RIS or fully-passive RIS architectures\cite{WuRISTutorial}. In semi-passive RIS, there are some active elements to enable sensing capability for channel estimation\cite{9053976,Jin_CE_semi-passiveRIS}. Most existing literature focuses on fully passive RIS due to its energy and cost efficiency. As seen in \cite{WuRISTutorial}, previous approaches often employed an ON/OFF method for sequential channel estimation by activating individual RIS elements one at a time. However, this method can become too complex and cause excessive training overhead due to the potential large number of RIS reflecting elements. A more effective approach was proposed in \cite{JensenOptimalCE_RIS}, where the channel estimation process is performed by exploiting the RIS reflection pattern with the aid of the discrete Fourier transform (DFT). There is an unbearable pilot overhead in RIS-assisted systems if we use conventional pilot-based channel estimation schemes due to the size of a RIS. The method in \cite{JensenOptimalCE_RIS} was extended by the authors in \cite{YangRIS_OFDM}, by introducing a transmission protocol that divides the RIS reflection elements into multiple groups. By estimating the channels for each group, the approach was able to reduce the training overhead. Compressed channel estimation for RIS was introduced by the authors in \cite{WangCompressedCE_RIS}, which exploits the inherent sparsity of mmWave channels to reduce overhead. Furthermore, the authors in \cite{Chen_CE_RIS_MIMO} proposed a novel channel estimation protocol based on compressive sensing techniques to estimate channels of the RIS-aided multi-user mmWave MIMO system. 

In a time-varying channel, channel coefficients vary according to Doppler shift changes due to user movements. In real-world scenarios, the channel may not change independently over time due to nearly stationary scattering environments. This concept is known as channel aging. The authors of \cite{JiangChannelAgingImpactRIS} theoretically analyzed the impact of channel aging on RIS-assisted systems. An investigation of the impact of channel aging on the performance of RIS-assisted massive MIMO systems under both spatial correlation and imperfect CSI conditions was presented in \cite{AnastasiosChannelAgingRIS}. The authors in \cite{ZhangChannelAgingPrecodingRIS} proposed an efficient suboptimal channel aging-aware precoding algorithm for RIS-aided multi-user communications.  All existing works show that channel aging can cause significant performance degradation in communication systems.  Moreover, channel aging can further increase the pilot overhead since more frequent estimations need to be carried out. The authors in \cite{Hu2TimeScaleCE_RIS} proposed a two-timescale estimation framework capable of reducing the pilot overhead by assuming that the channels between the BS and RIS are quasi-static and user-associated channels are time-varying. A three-stage joint channel decomposition and prediction framework based on the two-timescale property for a fast time-varying environment was proposed in \cite{XuTimeVaryChannelPredictionRIS}. Their prediction model was based on the long short-term memory (LSTM) neural network (NN) architecture. Both works assumed that the system operates under full-duplex mode. The authors of \cite{1512123}, examined the autoregressive (AR) modeling approach for the accurate prediction of correlated Rayleigh time-varying channels. A machine learning (ML)-based framework was proposed to improve the CSI prediction quality in \cite{YuanChannelAginML}. The authors of this work implemented a convolutional neural network (CNN) to identify the channel aging pattern and combined it with the AR method to predict the CSI in MIMO systems. 

The CNN-AR channel estimation method introduced in \cite{YuanChannelAginML} exhibited promising performance in terms of spectral efficiency and pilot overhead within standard MIMO systems. Nevertheless, the application of this method to RIS-assisted environments remains unexplored. This literature gap motivates the development of this work. In this paper, we extend the ideas of \cite{YuanChannelAginML} and propose a generalized CNN-AR framework for RIS-assisted MIMO systems. Specifically, a CNN model is designed and trained to identify the aging characteristics of the correlated time-varying wireless channels in a multi-user scenario. The obtained aging pattern is then integrated with the AR approach to forecast the desired CSI. Insightful simulation results are provided to assess the performance of the proposed model, revealing remarkable prediction accuracy. Our results also show that the achievable spectral efficiency of the implemented RIS-assisted MIMO system significantly improves when operating based on the predicted CSI. Furthermore, our proposed CNN-AR scheme demonstrates a reduced overhead compared to conventional channel estimation methods.

The rest of the paper is organized as follows. The system model and conventional AR model for channel prediction are described in Section \ref{SM}. In section \ref{CNN-AR}, we present the proposed ML-based channel prediction model. The numerical results are presented in Section \ref{results} and Section \ref{conclusion} concludes our paper.

\section{System Model}
\label{SM}
\begin{figure}[t]  
    \centering
    \includegraphics[width=0.45\textwidth]{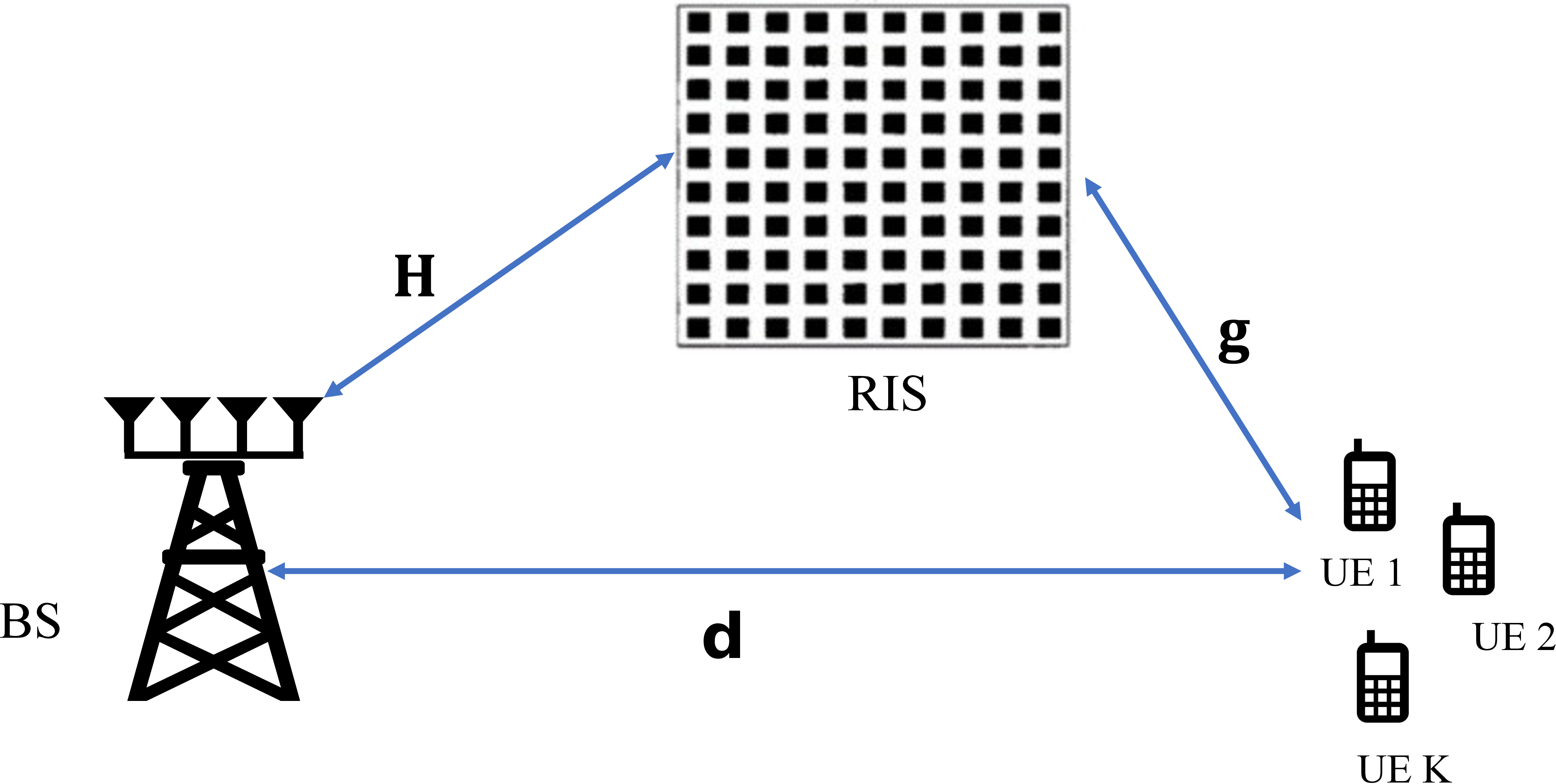}
    \caption{System model. An RIS is deployed to assist transmissions between a multi-antenna BS and multiple users.}
    \label{fig:Illustration of the system model}
    \vspace{-6mm}
\end{figure}

We consider an RIS-assisted MIMO system, as shown in Fig. \ref{fig:Illustration of the system model}, where a BS equipped with $N$ transmit antennas communicates with $K$ single-antenna user equipments (UEs).  An RIS composed of  $M \times M =\Tilde{M} $  reflecting elements is positioned between the BS and UEs. We partition the RIS into $M$ sub-surfaces by clustering neighboring elements that exhibit high correlation. Within each of these sub-surfaces, denoted by $m=1,\cdots,M$, we assume that the reflecting elements share a common coefficient. This grouping approach simplifies the representation of the RIS, allowing us to consider common characteristics within each sub-surface\cite{YangRIS_OFDM}.

The RIS is equipped with an intelligent controller that dynamically fine-tunes the phase shift of each element based on the instantaneous CSI obtained through periodic estimation. Specifically,  $\mathbf{d}_{k}[l] \in {\mathbb{C}^{N \times 1}}$ and $\mathbf{g}_k [l] \in \mathbb{C}^{M\times 1}$ denote, respectively, the BS-UE direct channel and the RIS-UE channel for the $k$th UE in $l$th coherence interval of duration $T_c$, and $\mathbf{H}\in {\mathbb{C}^{M \times N}}$ represents the BS-RIS channel.  Furthermore, we denote by ${\mathbf{G}}_k[l]= \mathbf{H}^H \mathrm{diag}(\mathbf{g}_k [l])\in \mathbb{C}^{N \times M} = [\mathbf{f}_{k,1}[l], \ldots, \mathbf{f}_{k,M}[l]]$ the cascaded channel associated with the $k$th UE in the $l$th coherence interval, without the effect of the RIS phase shifts. Generally,  channel aging does not affect the BS-RIS channel since the BS and RIS are usually fixed, making the channel quasi-static. Due to this reason, channel aging is modeled only in the BS-UE and RIS-UE channels. 

A total of $M$ distinct RIS coefficients need to be optimized as a result of the adopted RIS grouping strategy. More specifically, the coefficient associated with the $m$th sub-surface is expressed as $\phi_{m}=\alpha _{m} e^{j\theta _{m}}$, where $\theta _{m}\in [0,2\pi)$  is the  phase shift and $\alpha _{m}\in [{0,1}]$ is the reflection amplitude. In particular, we assume the ideal case that $\alpha _{m}=1$, $\forall m=1,2,\ldots, M$ to maximize the reflected power. The RIS coefficients are then organized into the following phase shift vector\cite{JiangChannelAgingImpactRIS} 
\begin{align}\bm{\theta} [l]=\left[\phi_1[l],\ldots, \phi_M[l] \right]^T.\end{align}
With the above definitions, the effective channel of the $k$th UE in the $l$th coherence interval can be written as
\begin{equation}\label{fullchannel}
{\mathbf{h}}_k[l] =  \mathbf{G}_k [l] \bm{\theta} [l]+ \mathbf{d}_{k}[l].
\end{equation}

We consider that the channel remains constant throughout each coherence interval but experiences variations from one interval to the next.  Moreover, the channels within a specific coherence interval are correlated with the channels in preceding intervals.

\subsection{Channel Estimation Scheme}

\begin{figure}[t]  
	\centering
	\includegraphics[width=0.48\textwidth]{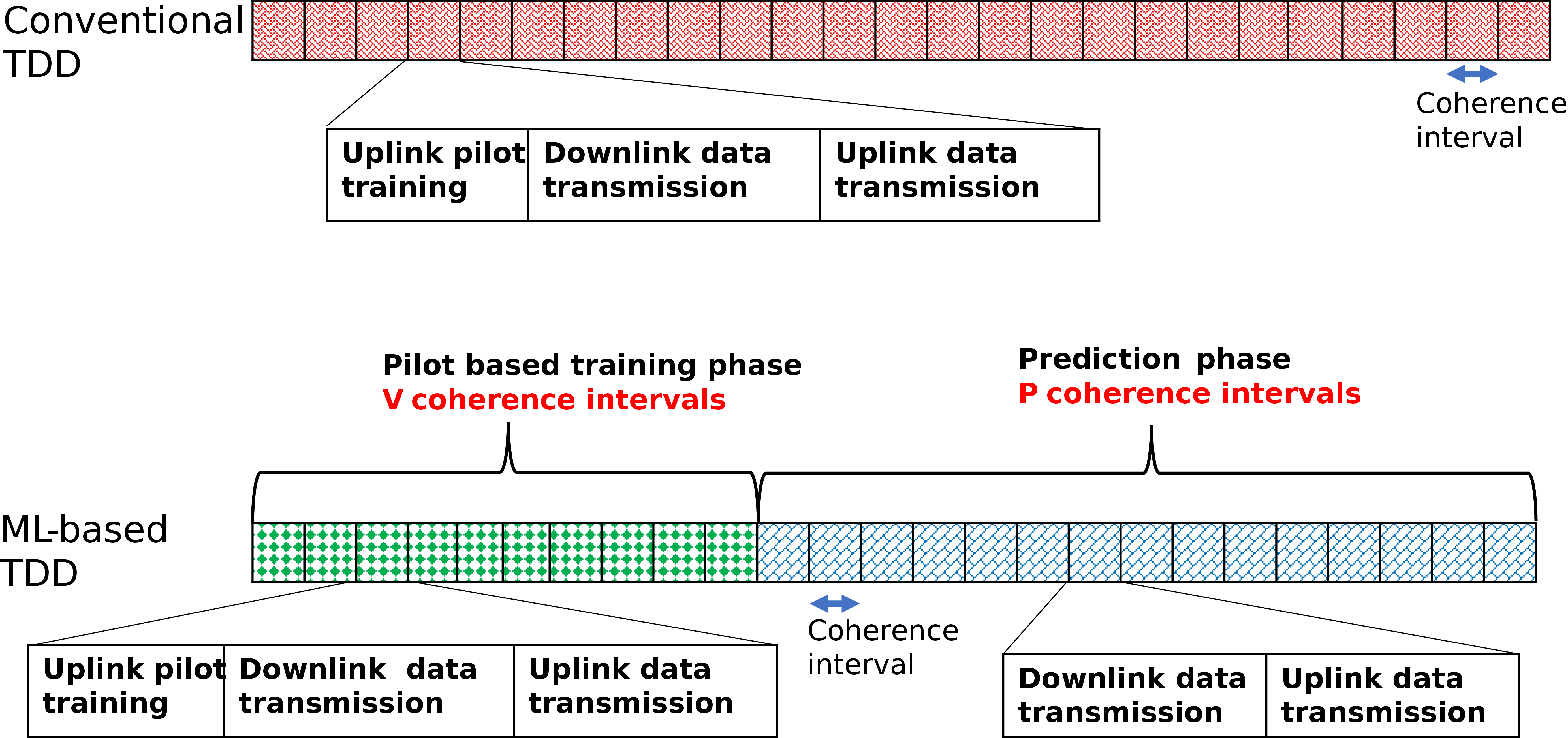}
	\caption{Simplified transmission block: Conventional TDD versus ML-based TDD. In the prediction phase, uplink pilot training is removed due to the introduction of ML-based CSI prediction.}
	\label{fig:Transmission Block}
 \vspace{-6mm}
 \end{figure}

We adopt a Time-Division Duplexing (TDD) protocol, which enables the use of the same frequency for both uplink and downlink transmissions, in which we leverage channel reciprocity for acquiring the initial CSI and subsequently predicting it. In our proposed ML-based channel prediction method, the transmission block is composed of a pilot-based training phase and a prediction phase, as shown in Fig. \ref{fig:Transmission Block}.

In conventional TDD-based approaches, within each coherence interval pilot symbols are sent in the uplink to estimate the direct and composite RIS channels. The estimated CSI is used for downlink and uplink data transmission by assuming that the channel is static during one coherence interval and that channel reciprocity is satisfied. In ML-based TDD, the pilot-based training phase, consisting of $V$ coherence intervals, works similarly as in conventional TDD. During the prediction phase, on the other hand, the channel estimates acquired in the first phase are exploited to predict the CSI through ML in the following $P$ coherence intervals. The predicted CSI is then used for downlink and uplink data transmission.   

Predefined reflection coefficients are used in the RIS during channel estimation, in which a DFT-based reflection matrix is employed \cite{JensenOptimalCE_RIS}. Moreover, orthogonal pilot sequences are sent by the users during the pilot-based training phase, which we denote by $\mathbf{x}_k [l] = [x_k [l, 1], \cdots, x_k [l, T]]$, such that $\mathbf{x}_k [l]^H \mathbf{x}_{k'} [l] = 0, \forall k \neq k'$, with $x_k [l, t]$ representing the pilot symbol transmitted in the $t$th time slot within the $l$th coherence interval, for $t = 1, \cdots, T$, with $T< T_c$. The received signal of the $k$th UE in $t$th time slot in the $l$th coherence interval is
\begin{equation} {\mathbf{y}_k}[l,t]= \sqrt{P_p}\left( {\mathbf{G}_k [l] \bm{\theta} [l, t]   + {\mathbf{d}}_{k}[l]} \right){{x}_k}[l,t]  + \mathbf{v}_k [l, t],\end{equation}
where ${P_p}$ is the power used by $k$th UE to transmit pilots. We then define:
\begin{align}\label{def1}
    \bm{\Theta}_k[l] &= \mathbf{X}_k[l]\bm{\Psi}[l],
\end{align}
and
\begin{align}\label{def2}
    \mathbf{\bar{y}}_k[l] = \hspace{-1mm}\left[ {\begin{array}{c} {{\mathbf{y}_k}[l,1]} \\ \vdots \\ {{\mathbf{y}_k}[l,T]} \end{array}} \right],\mathbf{\bar{v}}_k[l] = \hspace{-1mm}\left[ {\begin{array}{c} {{\mathbf{v}_k [l, 1]}} \\ \vdots \\ {{\mathbf{v}_k [l, T]}} \end{array}} \right],\mathbf{\bar{f}}_k[l] = \hspace{-1mm}\left[ {\begin{array}{c} \mathbf{d}_k[l] \\ \mathbf{f}_{k,1}[l] \\ \vdots \\ \mathbf{f}_{k,M}[l] \end{array}} \right],
\end{align}
where
\begin{align} \mathbf{X}_k[l] &= \operatorname{diag} \left( {\big[ {x_k [l, 1]{{1\!\!{\text l}}_{N\times1}}, \cdots ,x_k [l, T]{{1\!\!{\text l}}_{N\times1}}} \big]} \right), \notag
\end{align} 
\begin{align} \bm{\Psi}[l] = \bm{\Phi}[l] \otimes \mathbf{I}_N, \quad \bm{\Phi}[l] = \left[ {\begin{array}{cccc} 1&{{\phi _{1,1}[l]}}& \cdots &{{\phi _{1,M}[l]}} \\ \vdots & \vdots & \ddots & \vdots \\ 1&{{\phi _{T,1}[l]}}& \cdots &{{\phi _{T,M}[l]}} \end{array}} \right], \notag \end{align}
in which $\mathbf{\bar{v}}_k[l] \sim {\mathcal{C}}{\mathcal{N}}\left( {0,{\sigma_n^2}{\mathbf{I}_{N(M + 1)}}} \right)$ is the complex white Gaussian noise vector obtained during $T$ training periods.

By using the definitions in \eqref{def1} and \eqref{def2}, the pilot signals received from the $k$th UE during $T$ time slots within the $l$th coherence interval can be written as
%
\begin{align}
    \mathbf{\Bar{y}}_k[l] = \bm{\Theta}_k[l]\mathbf{\Bar{f}}_k[l] + \mathbf{\Bar{v}}_k[l].
\end{align}

It is noteworthy that our goal here is not to estimate $\mathbf{H}$ and $\mathbf{g}_{k}[l]$ separately since estimating the cascaded channel $\mathbf{G}_{k}[l]$ is sufficient for the objectives of this paper. This is achieved by employing a least square approach. More specifically, the least square estimation of the direct channel and the RIS-associated cascaded channel can be given by
\begin{align} \mathbf{\hat{f}}_k[l] = \left[ {\begin{array}{c} \mathbf{\hat{d}}_k[l] \\ \mathbf{\hat{f}}_{k,1}[l]  \\ \vdots \\ \mathbf{\hat{f}}_{k,M}[l] \end{array}} \right] &= \operatorname{argmin} ||\bm{\Theta}_k[l]\mathbf{\Bar{f}}_k[l]- \mathbf{\Bar{y}}_k[l]||_2^2 \notag \\ & = {\left( {{\bm{\Theta}_k[l]^H}\bm{\Theta}_k[l]} \right)^{ - 1}}{\bm{\Theta}_k[l]^H}\mathbf{\bar{y}}_k[l].\end{align} 

As can be seen, the estimate of the direct channel $\mathbf{\hat{d}}_k[l]$ can be retrieved from the first $N$ entries of $\mathbf{\hat{f}}_k[l]$, while the estimate of the cascade channel $\mathbf{\hat{G}}_k[l]$ can be obtained from the subsequent $M$ sub-vectors in $\mathbf{\hat{f}}_k[l]$. After obtaining these two channel components, the computation of the estimate $\mathbf{\hat{h}}_k[l]$ for the full channel in \eqref{fullchannel} is straightforward. This estimated CSI is used to find the optimal reflection coefficients for data transmission, as well as for the prediction of the channels in the prediction phase.

\subsection{Channel Aging Model}
The channel aging property primarily arises due to user mobility, and this characteristic can be approximately described using the second-order statistics of the channel, i.e., through the autocorrelation function (ACF)\cite{YuanChannelAginML}. The discrete-time ACF for the fading channel coefficients is given by
\begin{equation}R[l]=J_{0}(2\pi f_{n}\vert l\vert),\end{equation} where $J_{0}(\cdot)$ is the zeroth-order Bessel function of the first kind, in which $\vert l\vert$ is the delay in terms of the number of coherence intervals, $f_n$ is the normalized Doppler frequency defined as  $f_{n}=f_{\mathrm{d}}T_s$, normalized by the sampling rate $1/T_s$, and  $f_{\mathrm{d}}$ is the maximum Doppler frequency which is given by $f_{\mathrm{d}}=vf_{c}/c
$ , in which $v$ is the velocity of the $k$th UE, $c$ is the speed of light, and $f_{c}$ is the carrier frequency.

The AR stochastic model can be used to model channel aging phenomena, requiring only the channel correlation matching property \cite{YuanChannelAginML}. Specifically, we can model the channels of the considered RIS-assisted MIMO system as a small-scale correlated fading series as follows:
\begin{equation} \mathbf{h}_{k}[l]=-\sum\nolimits_{q=1}^{Q}a_{k,q}\mathbf{h}_{k}[l-q]+\boldsymbol{\omega}_k[l], \label{eq1} \end{equation}
where $\boldsymbol{\omega}_k[l]$ is an uncorrelated complex white Gaussian noise vector with zero mean and variance
\begin{equation} {\sigma_{\omega_k}}^{2}=R[0]+\sum\nolimits_{q=1}^{Q}a_{k,q}R[-q].
\label{eq2}\end{equation}
The AR coefficients $\{a_{k,q}\}_{q=1}^{Q}$ are evaluated via the Levinson-Durbin recursion as
\begin{equation}
    {\mathbf a} = - {\mathbf R}^{-1}{\mathbf w},
    \label{eq3}
\end{equation}
where 
\begin{equation}
 {\mathbf R} =\left[\begin{matrix}
R[0] & R[- 1] & \cdots & R[- Q + 1]\cr R[1] & R[0] & \cdots & R[- Q + 2]\cr\vdots & \vdots & \ddots & \vdots\cr R[Q - 1] & R[Q - 2] & \cdots & R[0] \end{matrix}\right],
\label{eq4}
\end{equation}
\begin{equation}
{\mathbf a} = [\begin{matrix} 
a_{k,1} & a_{k,2} & \cdots & a_{k,Q}\end{matrix}]^{T},
\label{eq5}
\end{equation}
\begin{equation}
 {\mathbf w} = \left[\begin{matrix} R[1] & R[2] & \cdots & R[Q]\end{matrix}\right]^{T}.
 \label{eq6}
\end{equation}

The parameter $Q$ in equations \eqref{eq1}--\eqref{eq6} represents the order of the AR model. The accuracy of the AR model improves with higher-order $Q$. The upper bound for $Q$ is equivalent to the number of coherence intervals used to collect CSI samples. 

Generating stable AR filters of high orders to accurately represent bandlimited channels can be challenging. A practical heuristic approach to address numerical issues was proposed in  \cite{1512123}. The authors presented a strategy to enhance the conditioning of the autocorrelation matrix ${\bf R}$ by slightly increasing the values in its main diagonal with a small positive amount $\epsilon$. By selecting a suitable $\epsilon$, the lower bound of $\sigma^2_{\omega_k}$ can be significantly increased, enabling the stable computation of larger-order AR models. Therefore, this strategy is adopted in this paper. Following \cite{1512123}, the first $Q+1$  ACFs of the resulting AR process can be reliably computed as  
\begin{equation}
\widehat{R}[q] = \begin{cases}
                   R[0] + \epsilon, & q = 0 \\
                   R[q], & q = 1,2,\ldots,Q.
                  \end{cases}
\end{equation}

\section{ML-Based Channel Prediction}
\label{CNN-AR}
\subsection{CNN-AR Model}

The performance of AR-based predictors generally improves with the increase of $Q$. The implication of this is that a simple AR predictor requires a large amount of CSI for high accuracy, which can be difficult to obtain in practice. Moreover, the BS-RIS-UE cascade channel is a combination of a time-invariant channel and a time-vary channel which makes it difficult to identify the channel aging pattern of the cascade channel compared to the BS-UE direct channel. To overcome these potential challenges, we propose a novel CNN-based prediction strategy to identify the channel aging pattern and determine accurate AR coefficients with arbitrary orders for a RIS-assisted MIMO system. 

\begin{figure}[t]  
	\centering
	\includegraphics[trim={0 15cm 0 0},scale=0.53]{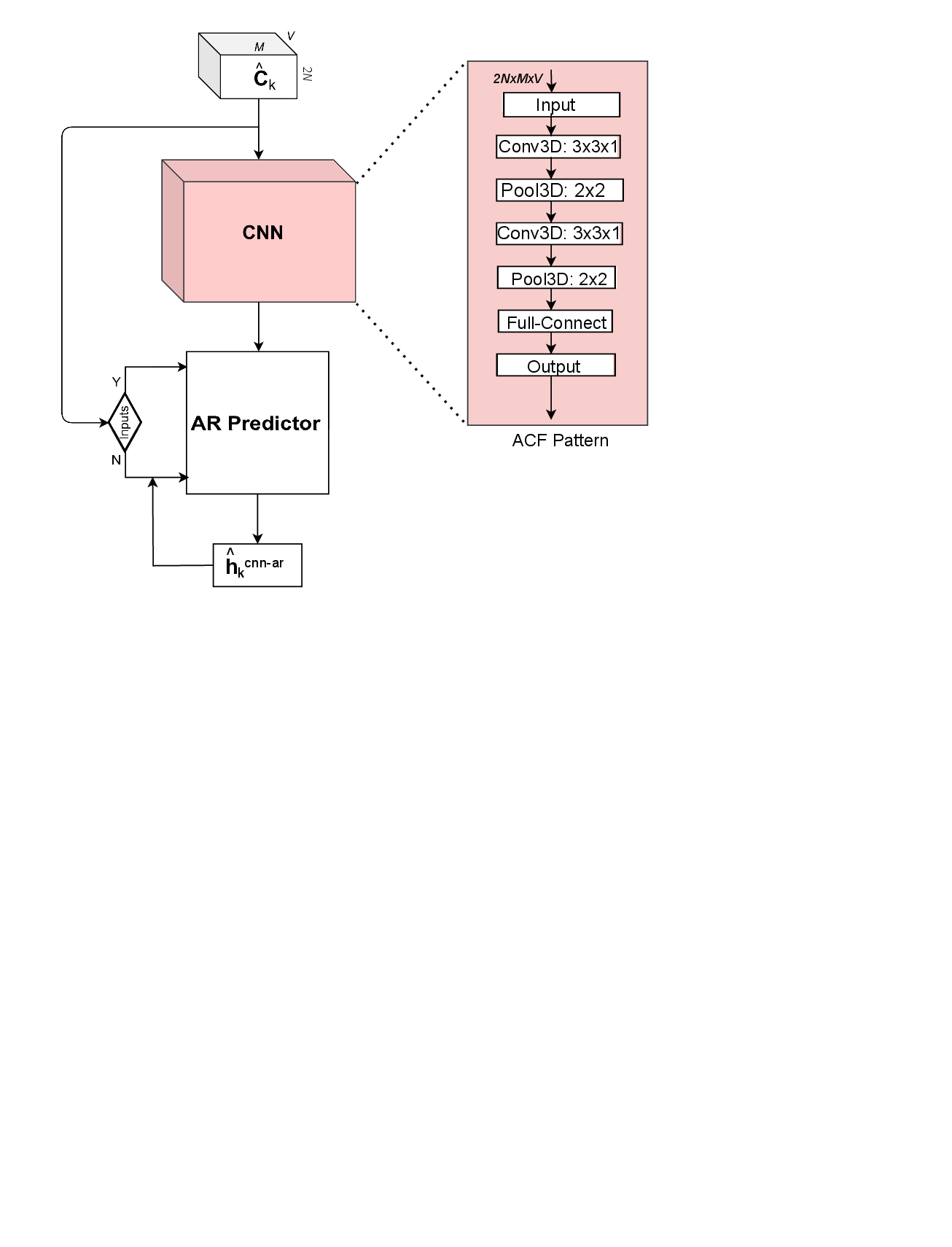}
	\caption{Architecture of the proposed CNN-AR predictor.}
	\label{fig:cnn-ar}
 \vspace{-6mm}
\end{figure}

 To this end, a CNN model is used to extract the ACF pattern by treating the CSI data as image data. The diagram of the proposed CNN-AR predictor is shown in Fig. \ref{fig:cnn-ar}. Our proposed CNN model can identify the ACF pattern in both the BS-UE direct channel and BS-RIS-UE cascaded channel.

The estimated channels $\mathbf{\hat{h}}_k[l]$, for $l=1,\cdots, V$, obtained during the pilot-based training phase are used as inputs for the implemented CNN model. This training data is first preprocessed, in which the estimated complex channel vectors collected during $V$ coherence intervals are transformed into a real-valued matrix, as follows:
\begin{align}
    \mathbf{\hat{C}}_k = \begin{bmatrix}
         \mathrm{Re}\{\mathbf{\hat{h}}_k[1]\} & \cdots & \mathrm{Re}\{\mathbf{\hat{h}}_k[V]\} \\
    \mathrm{Im}\{\mathbf{\hat{h}}_k[1]\} & \cdots & \mathrm{Im}\{\mathbf{\hat{h}}_k[V]\}
    \end{bmatrix}.
\end{align}

The proposed CNN model is used to identify the ACF pattern of the input CSI values. Subsequently, the proposed framework utilizes the AR coefficients associated with the aging pattern obtained by the CNN model. It then forecasts the CSI for the upcoming first coherence interval, as follows: (Y arrow direction for AR predictor in Fig. \ref{fig:cnn-ar})
\begin{equation} \mathbf{\hat{h}}_{k}^{\mathrm{cnn-ar}}\left [{ l }\right] = - \sum \nolimits _{q = 1}^{Q} {a_{k,q}} \mathbf{\hat{h}}_k\left[{l - q}\right].\end{equation}

The result for the current coherence interval is used to predict the CSI for the next $P-1$ intervals as follows:  (N arrow direction for AR predictor in Fig. \ref{fig:cnn-ar})
\begin{align}&\hspace {-2pc}\mathbf{\hat{h}}_{k}^{{\mathrm{cnn-ar}}} \left [{ {l + l'} }\right] \notag \\=&- \sum \nolimits _{q = l' + 1}^{Q} {a_{k,q}{\mathbf{\hat{h}}}_{k}\left [{ {l + l' - q} }\right]} \notag \\&-\, \sum \nolimits _{q' = 1}^{l'} {a_{k,q'}}{\mathbf{\hat{h}}_{k}^{{\mathrm{cnn-ar}}}\left [{ {l + l' - q'} }\right]},\;\;\; l'\in P.\end{align}

As for the model architecture, we utilize the tanh activation function in the convolutional layers and sigmoid in fully connected layers. We implement, specifically, two fully connected layers with $512$ and $256$ neurons. Furthermore, we adopt the adaptive moment estimation (Adam) optimizer and employ the mean-squared error (MSE) loss function. 

\subsection{Pilot overhead}
The channel estimation of uplink channels demands in general a pilot overhead of $P_O^{\mathrm{conv}} = (\Tilde{M}NK+NK)(V+P)$ if we are using conventional TDD channel estimation methods, as demonstrated in \cite{JensenOptimalCE_RIS}. On the other hand, our proposed channel estimation scheme only demands $P_O^{\mathrm{cnn-ar}}=(MNK+NK)V$ pilots during $V+P$ coherence intervals. Numerically, we can show that $P_O^{\mathrm{conv}} \gg P_O^{\mathrm{cnn-ar}}$ since $\Tilde{M} \gg M$ and $V+P \geq V$. Therefore, our proposed channel estimation scheme has low pilot overhead compared to conventional TDD schemes.

\subsection{Performance Metrics}
We adopt two important metrics to analyze the performance of our proposed scheme, which are explained in the following.

\subsubsection{Prediction Accuracy}
 First, we use the normalized mean square error (NMSE) to evaluate the accuracy of the prediction. The NMSE metric is computed as
 \begin{align} \mathrm{NMSE}\left [{l}\right]={\mathrm {E}}\left \{ \frac {1}{K}\sum_{k = 1}^{K}
 \frac{ \left \|{\mathbf{\hat{h}}_{k}^{{\mathrm{cnn-ar}}} \left [{ l }\right] - {{\mathbf{{h}}}_{k}}\left [{ l }\right]} \right \|_{2}^{2}}{ {{ {\left \|{ {{{\mathbf{{h}}}_{k}}\left [{ l }\right]} }\right \|_{2}^{2}}}} } \right \}.\end{align}

\subsubsection{Spectral Efficiency}
The second performance metric that we investigate is the average spectral efficiency of the system. To this end, we jointly optimize the active beamforming vector for the $k$th UE, denoted as $\mathbf{u}_k[l]$, and the passive reflection coefficients $\bm{\theta} [l]$ of the RIS, aiming at maximizing the spectral efficiency during the downlink data transmission phase. The spectral efficiency is defined as
\begin{align}
\mathrm{SE}_k[l]=\log _{2}\left({1+\frac {P_{d}}{\sigma _{n}^{2}}\left |{(\mathbf{G}_k [l] \bm{\theta} [l] +\mathbf {d}_{k}[l])^H\mathbf {u}_k}[l]\right|^{2}}\right),
\end{align}
where $P_d$ is the power used to transmit a data symbol and $\sigma_n^2$ is the noise variance. As a result, our objective can be achieved by solving the following optimization problem:
\begin{align}(\mathrm{P}1):&\max _{{\theta}_m[l] ,\:\mathbf {u}_k[l]}\quad \biggl |\Bigl(\mathbf{G}_k [l] \bm{\theta} [l] +\mathbf {d}_{k}[l])\Bigr)^H\mathbf {u}_k[l]\biggr |^{2} \notag \\&\textrm {s.t.} \quad \|\mathbf {u}_k[l]\|^{2}\leqslant 1 \notag \\&\hphantom {\textrm {s.t.} \quad } \theta _{m}[l]\in [0,2\pi), \quad \forall m=1,2,\ldots,M. \label{opt1} \end{align}

The optimization problem (P1) is non-convex. We address this challenge by employing an alternating optimization approach. Initially, we keep the passive reflection coefficients $\bm{\theta} [l]$ fixed and find the optimal beamforming vector $\mathbf{u}_k[l]$, which is given by
\begin{equation}
    \mathbf {u}^{\star}_k[l] = \frac {\left (\mathbf{G}_k [l] \bm{\theta} [l]  +\mathbf {d}_{k}[l])\right)^{H}}{\|\mathbf{G}_k [l] \bm{\theta} [l]  +\mathbf {d}_{k}[l])\|}.
\end{equation}

 By substituting $\mathbf {u}^{\star}_k[l] $ to (P1) it can be simplified as the following equivalent problem.
\begin{align} (\mathrm{P}2):\quad \max_{\theta_m[l]}\qquad& \Vert (\mathbf{G}_k [l] \bm{\theta} [l]   +\mathbf {d}_{k}[l])\Vert^{2} \notag \\ \mathrm{s}.\mathrm{t}.\qquad&\ 0\leq\theta_{m}[l]\leq 2\pi,\ \forall m=1,\cdots,M.  \end{align}
(P2) is solved to find optimal reflection coefficients, using an algorithm based on the semidefinite relaxation (SDR) technique as performed in \cite{WuBeamformingRIS}. The average spectral efficiency of $k$th UE is $\bar{\mathrm{SE}}_k[l] =\frac {1}{N_s}\sum _{l=1}^{N_s} \mathrm{SE}_k[l]$ where $N_s$ is the total coherence intervals.

\begin{figure}[t]  
	\centering
	\includegraphics[width=0.3\textwidth]{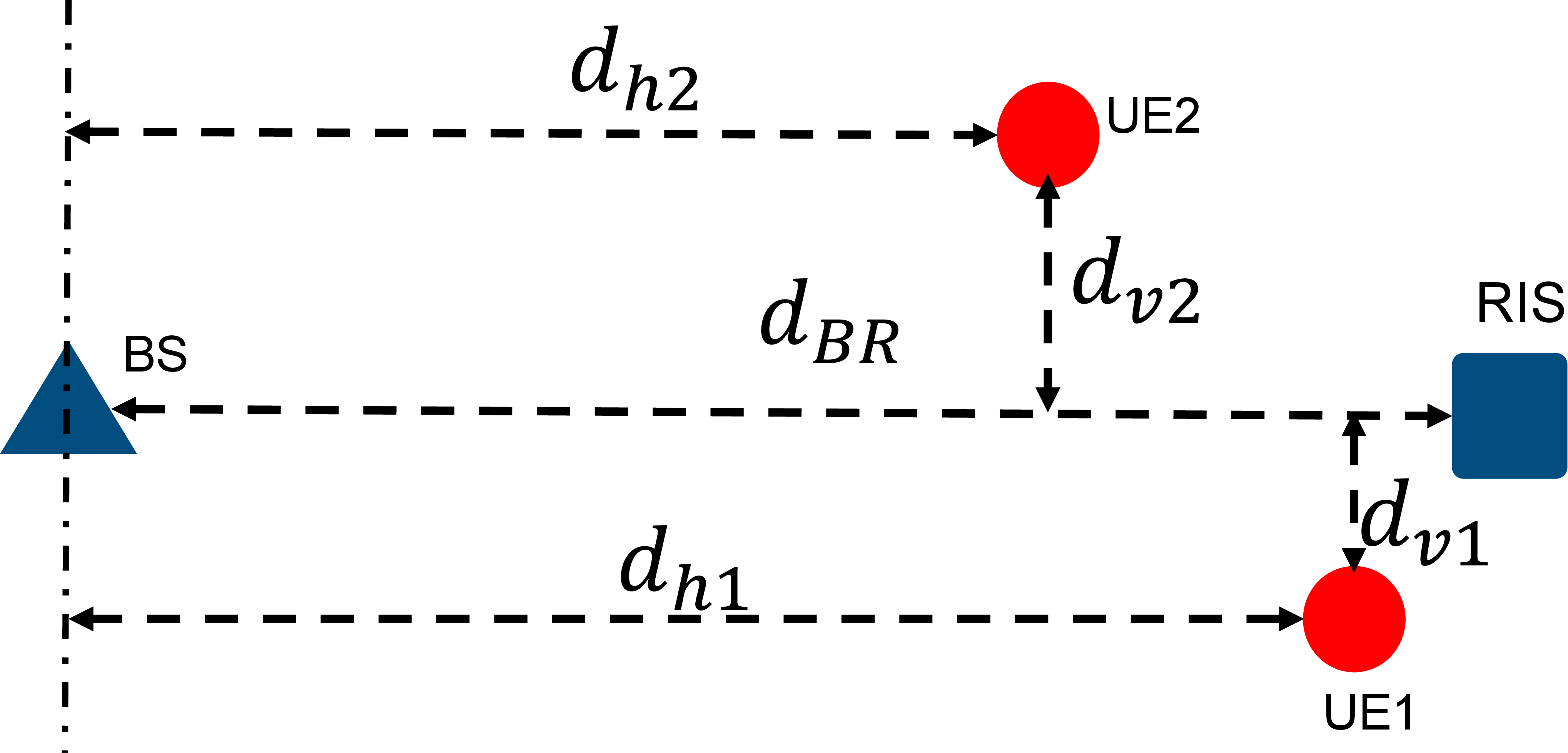}
	\caption{ Simulation setup. Scenario with two users.}
	\label{fig:Simulation setup}
 \vspace{-4mm}
\end{figure} 

\section{Numerical Results }
\label{results}
In this section, we present insightful simulation results to assess the performance of our proposed CNN-AR model. The considered geometrical scenario is illustrated in Fig. \ref{fig:Simulation setup} for the case with $2$ UEs. Furthermore, we assume $d_{BR}=51$~m, while various values for the other distances are considered in our simulations. Following the model in Fig. \ref{fig:Simulation setup}, the BS-UE and RIS-UE distances for the $k$th UE are computed by $d_{\text{BU} k}=\sqrt {d_{hk}^{2}+d_{vk}^{2}}$ and $d_{\text{RU} k}=\sqrt {(d_{BR}-d_{hk})^{2}+d_{vk}^{2}}$. The large-scale fading is molded as
\begin{equation}
L(d)=L_{0}/d^{-\alpha },
\end{equation}
where $L_0$ is the path loss at the reference distance of $1$~m, $d$ is the distance, and $\alpha$ is the path-loss exponent. We also model path loss due to shadowing, which is set to $10$~dB and is applied to both BS-UE and RIS-UE links. Other simulation parameters are listed in Table \ref{table:simulation parameters}. The training, validation, and testing datasets have been generated with $7\times 10^4$, $2\times 10^4$, and $1\times 10^4$ samples, respectively, for the prediction phase. Moreover, we considered $7000$, $2000$, and $1000$ UEs for the training, validation, and testing datasets, respectively, with $10$ different $f_d$ values and various distance values. The ML model undergoes $300$ epochs while the early-stopping is employed to avoid overfitting during the training phase, with a batch size of $50$, while the learning rate is fixed at $0.001$. 

\begin{table}[h]
	\caption{Simulation Parameters} 
	\centering 
	\begin{tabular}{|c| c| } 
		\hline 
	    $N$ & $12$\\
		\hline
		$\Tilde{M}$ & $225$\\
	    \hline
		$L_0$&-$30$~dB\\
		\hline
            $P_p$ ,$P_d$ & $0$~dBm,  $5$~dbm\\
            \hline
            $\sigma^2_n$ & $-174$~dBm \\
            \hline
            $f_c$ & $3$~GHz \\
            \hline 
            $\epsilon$ & $0.1$ \\
            \hline
            $\alpha$ for BS-UE,  RIS-UE , and BS-RIS links & $3, 3$, and $2$ \\
            \hline
	\end{tabular}
	\label{table:simulation parameters} 
\end{table}

\begin{figure}[t]  
	\centering
	\includegraphics[width=0.48\textwidth]{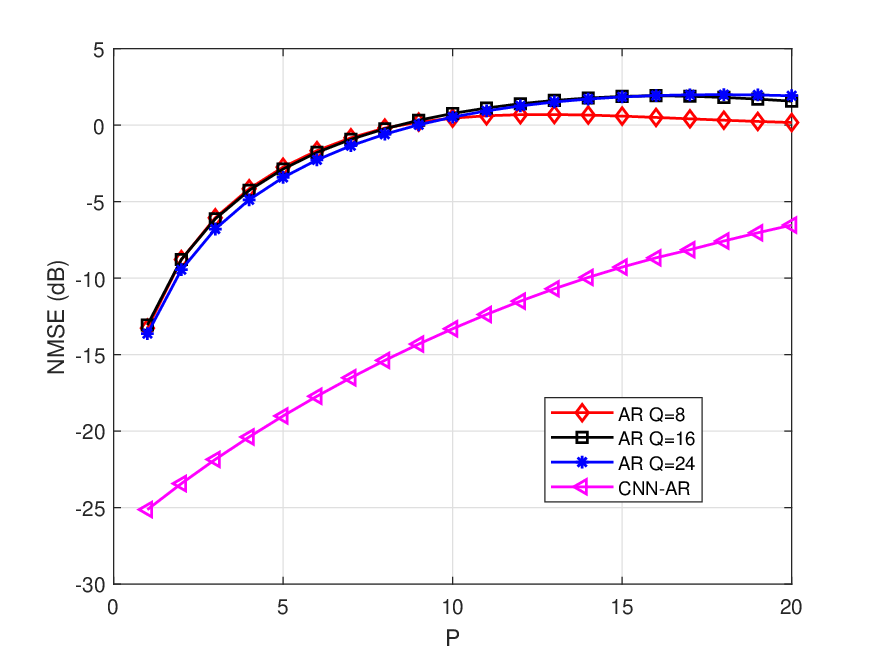}
	\caption{ Comparison of prediction NMSE among the AR predictors and CNN-AR for the first $P$ intervals.}
	\label{fig:cnn-ar comparison RIS}
 \vspace{-4mm}
\end{figure}

The NMSE is chosen to evaluate the prediction performance. Fig. \ref{fig:cnn-ar comparison RIS} shows the comparison of prediction NMSE among the AR predictors, and CNN-AR for the first $P$ intervals. 
According to Fig. \ref{fig:cnn-ar comparison RIS}, the CNN-AR predictor has high prediction accuracy compared to AR estimators. In this simulation, we assume $f_d=50$~Hz ($18$~kmph). Moreover, according to  Fig. \ref{fig:cnn-ar comparison RIS}, comparison between $Q=8,16$ and $24$ shows that by expanding the order, the prediction accuracy is hardly improved. The reason behind the huge improvement in the prediction accuracy of the CNN-AR predictor is that the real-time calculation of the AR coefficient based on a limited number of CSI data inputs is not accurate enough. In contrast, the CNN-AR model can find the channel variation pattern and load the pre-computed AR coefficients in arbitrary order.

\begin{figure}[t]
    \centering
    \includegraphics[width=0.47\textwidth]{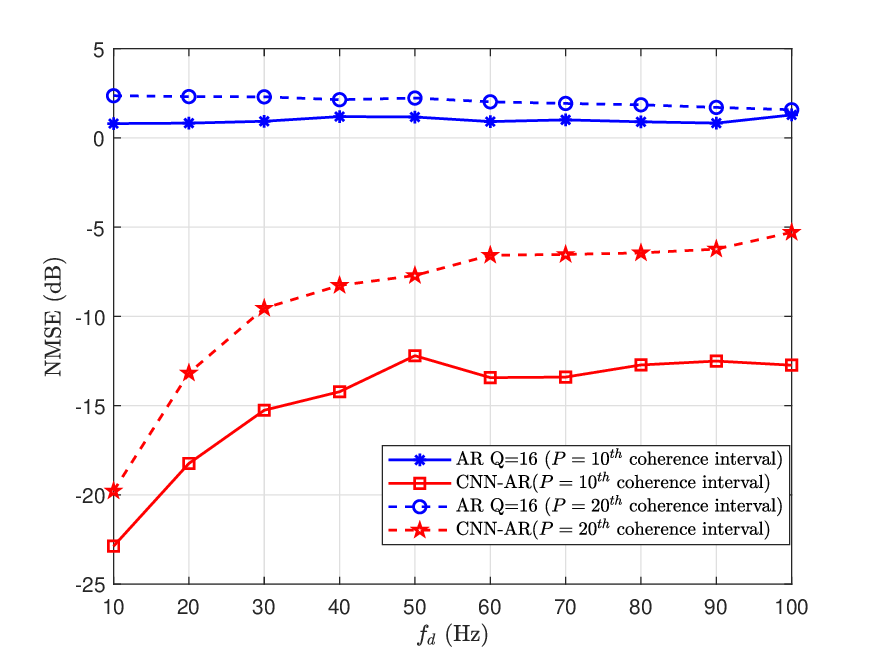}
    \caption{Comparison of prediction NMSE among the AR predictor and CNN-AR for different $f_d$ in $P=10^{th}$ and $20^{th}$ coherence intervals.}
    \vspace{-5mm}
	\label{fig:fd}
\end{figure}
\begin{figure}[t]  
	\centering	\includegraphics[width=0.47\textwidth]{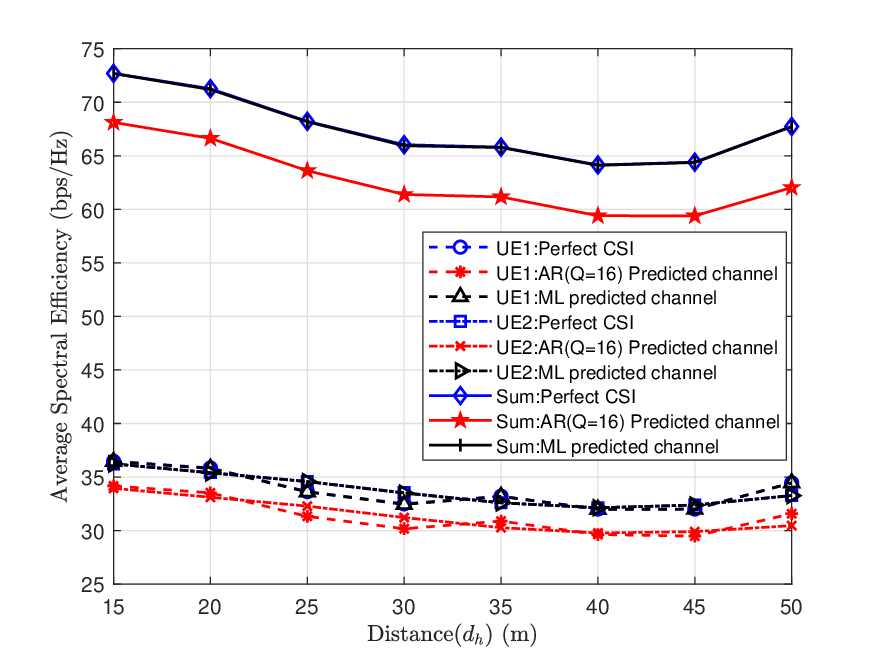}
	\caption{Average spectral efficiency for the scenario with two UEs versus different values of $d_h$, for $d_{h} = d_{h1} = d_{h2}$.}
	\label{fig:se comparison RIS}
    \vspace{-5mm}
\end{figure}

Fig. \ref{fig:fd} shows the comparison of prediction NMSE among the AR predictor and CNN-AR for different Doppler shift values in $P=10^{th}$ and $20^{th}$ coherence intervals. NMSE is increasing for prediction in later coherence intervals in both the AR-predictor and CNN-AR model. Even though, the CNN-AR model has a lower NMSE at the $20^{th}$ coherence interval compared to the NMSE of the AR-predictor's $10^{th}$ coherence interval.  The range of the $v$ varies from  $3.6$~kmph to $36$~kmph, correspondingly $f_d$ varies from $10$~Hz to $100$~Hz. The comparison of the average spectral efficiency of the AR estimator and CNN-AR is shown in Fig. \ref{fig:se comparison RIS}. In this simulation we consider 2 UEs as in Fig. \ref{fig:Simulation setup} with $d_h=d_{h1}=d_{h2}$, $d_{v1}=2$~m and $d_{v2}=3$~m. It shows that it has a higher average spectral efficiency when channels are predicted using the CNN-AR model compared to the AR estimator. Furthermore, it shows that the average spectral efficiency when channels are predicted using the CNN-AR model is similar to the perfect CSI scenario.  Moreover, Fig. \ref{fig:se comparison RIS} shows that average spectral efficiency is higher when the UE lies near either the BS or RIS, due to the reduction in the ``double fading'' effect of the cascaded RIS channel.

\section{Conclusion}
\label{conclusion}
In this paper, we proposed an efficient CNN-AR-based solution to address the challenges associated with channel estimation in correlated multi-user RIS-assisted MIMO systems. Specifically, through the integration of a CNN model with an AR predictor, our proposed extended channel estimation framework demonstrated robustness to channel aging in correlated fast-fading environments. The implemented CNN-AR approach successfully identified aging patterns and provided enhanced estimates of wireless channels, showcasing its high-precision estimation accuracy. Moreover, our insightful simulation results highlighted the superiority of our approach in achieving high spectral efficiency and low pilot overhead when compared to traditional methods.

\bibliographystyle{IEEEtran}
\bibliography{IEEEabrv,main}{}

\begin{thebibliography}{10}
\providecommand{\url}[1]{#1}
\csname url@samestyle\endcsname
\providecommand{\newblock}{\relax}
\providecommand{\bibinfo}[2]{#2}
\providecommand{\BIBentrySTDinterwordspacing}{\spaceskip=0pt\relax}
\providecommand{\BIBentryALTinterwordstretchfactor}{4}
\providecommand{\BIBentryALTinterwordspacing}{\spaceskip=\fontdimen2\font plus
\BIBentryALTinterwordstretchfactor\fontdimen3\font minus \fontdimen4\font\relax}
\providecommand{\BIBforeignlanguage}[2]{{%
\expandafter\ifx\csname l@#1\endcsname\relax
\typeout{** WARNING: IEEEtran.bst: No hyphenation pattern has been}%
\typeout{** loaded for the language `#1'. Using the pattern for}%
\typeout{** the default language instead.}%
\else
\language=\csname l@#1\endcsname
\fi
#2}}
\providecommand{\BIBdecl}{\relax}
\BIBdecl

\bibitem{Liu21_RIS_survey}
Y.~Liu \emph{et~al.}, ``{Reconfigurable Intelligent Surfaces: Principles and Opportunities},'' \emph{IEEE Commun. Surv. Tutorials}, vol.~23, no.~3, pp. 1546--1577, 2021.

\bibitem{WuRISTutorial}
Q.~{Wu}\'{\i}n \emph{et~al.}, ``{Intelligent Reflecting Surface Aided Wireless Communications: A Tutorial},'' \emph{IEEE Trans. Commun.}, pp. 1--1, 2021.

\bibitem{9053976}
G.~C. Alexandropoulos and E.~Vlachos, ``{A Hardware Architecture For Reconfigurable Intelligent Surfaces with Minimal Active Elements for Explicit Channel Estimation},'' in \emph{ICASSP}, 2020, pp. 9175--9179.

\bibitem{Jin_CE_semi-passiveRIS}
Y.~Jin \emph{et~al.}, ``{Channel Estimation for Semi-Passive Reconfigurable Intelligent Surfaces With Enhanced Deep Residual Networks},'' \emph{IEEE Trans. Veh. Technol.}, vol.~70, no.~10, pp. 11\,083--11\,088, 2021.

\bibitem{JensenOptimalCE_RIS}
T.~L. {Jensen} and E.~{De Carvalho}, ``{An Optimal Channel Estimation Scheme for Intelligent Reflecting Surfaces Based on a Minimum Variance Unbiased Estimator},'' in \emph{ICASSP}, 2020, pp. 5000--5004.

\bibitem{YangRIS_OFDM}
Y.~Yang \emph{et~al.}, ``{Intelligent Reflecting Surface Meets OFDM: Protocol Design and Rate Maximization},'' \emph{IEEE Trans. Commun.}, vol.~68, no.~7, pp. 4522--4535, 2020.

\bibitem{WangCompressedCE_RIS}
P.~Wang \emph{et~al.}, ``{Compressed Channel Estimation for Intelligent Reflecting Surface-Assisted Millimeter Wave Systems},'' \emph{IEEE Signal Process Lett.}, vol.~27, pp. 905--909, 2020.

\bibitem{Chen_CE_RIS_MIMO}
J.~Chen \emph{et~al.}, ``{Channel Estimation for Reconfigurable Intelligent Surface Aided Multi-User mmWave MIMO Systems},'' \emph{IEEE Trans. Wireless Commun.}, vol.~22, no.~10, pp. 6853--6869, 2023.

\bibitem{JiangChannelAgingImpactRIS}
W.~Jiang and H.~D. Schotten, ``{Performance Impact of Channel Aging and Phase Noise on Intelligent Reflecting Surface},'' \emph{IEEE Commun. Lett.}, vol.~27, no.~1, pp. 347--351, 2023.

\bibitem{AnastasiosChannelAgingRIS}
A.~Papazafeiropoulos, I.~Krikidis, and P.~Kourtessis, ``{Impact of Channel Aging on Reconfigurable Intelligent Surface Aided Massive MIMO Systems With Statistical CSI},'' \emph{IEEE Trans. Veh. Technol.}, vol.~72, no.~1, pp. 689--703, 2023.

\bibitem{ZhangChannelAgingPrecodingRIS}
Y.~Zhang \emph{et~al.}, ``{Channel Aging-Aware Precoding for RIS-Aided Multi-User Communications},'' \emph{IEEE Trans. Veh. Technol.}, vol.~72, no.~2, pp. 1997--2008, 2023.

\bibitem{Hu2TimeScaleCE_RIS}
C.~Hu \emph{et~al.}, ``{Two-Timescale Channel Estimation for Reconfigurable Intelligent Surface Aided Wireless Communications},'' \emph{IEEE Trans. Commun.}, vol.~69, no.~11, pp. 7736--7747, 2021.

\bibitem{XuTimeVaryChannelPredictionRIS}
W.~Xu \emph{et~al.}, ``{Time-Varying Channel Prediction for RIS-Assisted MU-MISO Networks via Deep Learning},'' \emph{IEEE Transactions on Cognitive Communications and Networking}, vol.~8, no.~4, pp. 1802--1815, 2022.

\bibitem{1512123}
K.~Baddour and N.~Beaulieu, ``{Autoregressive modeling for fading channel simulation},'' \emph{IEEE Trans. Wireless Commun.}, vol.~4, no.~4, pp. 1650--1662, 2005.

\bibitem{YuanChannelAginML}
J.~Yuan, H.~Q. Ngo, and M.~Matthaiou, ``{Machine Learning-Based Channel Prediction in Massive MIMO With Channel Aging},'' \emph{IEEE Trans. Wireless Commun.}, vol.~19, no.~5, pp. 2960--2973, 2020.

\bibitem{WuBeamformingRIS}
Q.~Wu and R.~Zhang, ``{Intelligent Reflecting Surface Enhanced Wireless Network: Joint Active and Passive Beamforming Design},'' in \emph{IEEE GLOBECOM}, 2018, pp. 1--6.

\end{thebibliography}

\end{document}